\documentclass[12pt]{JHEP3}
\usepackage{pb-diagram,lamsarrow,pb-lams,amsfonts,amssymb,amsthm}
\usepackage{graphicx}
\usepackage{cite}
\usepackage{amsmath,epsfig}
\def\be{\begin{equation}}
\def\ee{\end{equation}}
\def\bea{\begin{aligned}}
\def\eea{\end{aligned}}
\def\ba{\begin{eqnarray}}
\def\ea{\end{eqnarray}}

\def\be{\begin{equation}}
\def\ee{\end{equation}}
\def\bea{\begin{eqnarray}}
\def\eea{\end{eqnarray}}

\def\yzero{\smash{\hbox{$y\kern-4pt\raise1pt\hbox{${}^\circ$}$}}}

\def\beq{\begin{equation}}
\def\eeq{\end{equation}}
\def\beqa{\begin{eqnarray}}
\def\eeqa{\end{eqnarray}}

\def\-{\hphantom{-}}

\def\s2{\frac{1}{\sqrt2}}

\def\beq{\begin{equation}}
\def\eeq{\end{equation}}
\def\beqa{\begin{eqnarray}}
\def\eeqa{\end{eqnarray}}

\def\IF{\relax{\rm I\kern-.18em F}}
\def\II{\relax{\rm I\kern-.18em I}}
\def\IP{\relax{\rm I\kern-.18em P}}
\def\IC{\relax\hbox{\kern.25em$\inbar\kern-.3em{\rm C}$}}
\def\IR{\relax{\rm I\kern-.18em R}}

\def\Dsl{\,\raise.15ex\hbox{/}\mkern-13.5mu D} 
\def\IZ{Z\kern-.4em  Z}

\usepackage{cite}

\title{   Axion   from  Quivers in Type II Superstrings}

\author{Adil Belhaj$^{1}$, Salah Eddine Ennadifi$^{2}$, Maria Pilar Garcia del Moral$^{3}$\footnote{E-mail:
\emph{ belhaj@unizar.es,  ennadifis@gmail.com, maria.garciadelmoral@uantof.cl}}\\
$^1${D\'epartement de Physique, LSIRT,  Facult\'e Polydisciplinaire,
Universit\'e Sultan Moulay Slimane, B\'eni Mellal, Morocco\\
 $^2$Laboratoire des Physiques des Hautes \'{e}nergies: Mod\'elisation
et simulation, Universit\'e Mohammed V,Rabat, Morocco\\
$^3$Departamento de F\'isica, Universidad de Antofagasta,
Antofagasta, Aptdo 02800, Chile}}

\abstract{  We  investigate a string-inspired axion extension of the
standard model obtained from  Type II superstrings   using quiver
method. In the first
 part,  we discuss  intersecting Type IIA D6-branes wrapping non
trivial 3-cycles in the presence of the Peccei-Quinn symmetry U(1)$_{PQ}$.
Concretely, a complex scalar field $\phi =\rho exp(\frac{i\sigma}{%
f_{\sigma}})$, where $\sigma$ is a closed string axion generates a
general fermion Yukawa coupling weighted by a flavor-dependent power
$n_{f}$ taking specific values. Using string theory and standard
model data,  we find  that the corresponding  axion window is in the
allowed range $10^{9}GeV\leq f_{\sigma}\leq 10^{12}GeV$ matching
with the recent cosmological results.  Then,  we  extend these
results  to  the case of  the hyperbolic quiver  whose the moduli is
related to the stringy axion using root systems of  �ADE  Lie
algebras. For the hyperbolic quiver case,  we observe that the
closed axion decay constant becomes disentangled from the string
scale.}

\keywords{Type II Superstrings, Axion, hyperbolic quivers, D-branes
and Standard Model, ADE Lie algebras} \voffset=-1.35in

\date{8/6/15}
\begin{document}
\section{Introduction}
Low-energy phenomena have been successfully described by the
Standard Model (SM) of particle physics. With the recent observation
of the pioneering piece, i.e. the Higgs boson, a large amount of
experimental data is now said to be in an excellent agreement with
SM physics\cite{1,2,3}. However, some famous issues mostly related
to the hierarchy problem, fermion masses and dark matter are still
considered as open questions \cite{4,5}. In particular, many stringy
inspired models have been elaborated exploring either gauge theories
obtained from intersecting Type II D-branes wrapping non trivial
cycles in  Calabi-Yau manifolds  or M-theory compactified on $G_2$
manifolds \cite{05,005,6,7,70,71,8}. In string theory approach, it has
been argued that open strings stretched to different stacks of Type
II D-branes
produce the SM particle spectrum. More precisely, a unitary group $\mbox{U}%
(N)\sim \mbox{SU}(N)\times \mbox{U}(1)$ is generated by $N$
coincident Type II D-branes and, typically, the gauge group model
gets extended with extra abelian factors \cite{9,10}. Generically,
those abelian fields have four-dimensional anomalies canceled via
the Green-Schwarz mechanism which gives a mass to the anomalous U(1)
fields by breaking the associated gauge symmetry \cite{11,6,14}.
This provides an acceptable effective low-energy description
reproducing either SM-like models or more generally some extensions
\cite{15,16}. In these models, the gauge and the matter fields could
be encoded in quivers representing stacks of coinciding Type II
D-branes. Working at the level of quivers one could investigate the
Yukawa-like couplings as well as their strengths from the quiver
data rather than examining  the whole geometrical and physical data
derived from string theory \cite{17,18,19}.

Recently, interesting Type II superstring inspired models in the
presence of various types of axionic fields have been investigated
using different approaches. It has been shown that such fields play
a primordial role in the physics beyond the SM and cosmology
\cite{20,21,22,220}. In particular, globally consistent D6-brane
models with SM spectrum in the Type IIA orientifold compactification
scenario have been elaborated. Different energy scales have been
discussed including the Peccei-Quinn U(1)$_{PQ}$ symmetry associated
with electroweak and supersymmetry breaking scales in the context of
closed and open string sectors \cite{230,231,23}.  Related  issue
related to axions in QCD and inflation have been  extensively studied in
\cite{371,28,280,281,fgu,290,30,311,31}.

 In the study of stringy phenomenological models,
hyperbolic geometries  are not so much used. However,
  recently there has been several attempts in cosmology,
    to realize inflation through axions associated
to hyperbolic geometries \cite{31,311}.   It has been studied M-theory
solutions involving compact hyperbolic
spaces\cite{312,313,314,315,316}. These spaces allow for
accelerating cosmologies and Randall-Sundrum realization, many of
them break completely supersymmetry and do not require light Kaluza
Klein fields. All of them  are considered relevant aspects for
the particle physics phenomenology. On the other hand, hyperbolic
spaces are characterized in principle by less number of moduli
(which is a virtue in order to simplify the stabilization of the
moduli). The only relevant length scales of the hyperbolic spaces
are a characteristic length $l$ associated with  the local
properties of the concrete model as for example the curvature of the compact space, and a second one $L$ related  to
the global aspects of those spaces, such as the volume of
the nontrivial cycle considered. These two scales usually
disentangle and this has allowed in the past to explain the 4D
Planck and String hierarchy of scales. A very nontrivial fact that
we will use in our analysis. Examples of this constructions are
geometries of the type $AdS_{7-d}\times H_d/\Gamma\times S^4$ for
$d>2$ which  are associated with solutions denoted by 'Swiss Cheese
Universe' obtained from 11D supergravity or the uplift of Lifshitz geometries to string theory\cite{erika} $Li\times \Omega$ with $\Omega$ some hyperbolic manifold. Some of the hyperbolic
spaces geometries often produce de Sitter spaces with no
supersymmetry \cite{313} and also they are useful to explain
late-time cosmologies.

 Other
relevant property  is the
 deep connection between supergravity and hyperbolic Kack-Moody symmetries
 of infinite type $K(E_{10})$ \cite{33p}, and Standard model matter content, as signaled in  \cite{33p,34}.  Recently,  it has been
   possible to identify the spin half integer fermions of the gauged $SO(8)$
    supergravity with the matter content of the Standard Model, providing
    a new way to embed Standard model physics into the low energy limit a quantum
     gravity theory  \cite{33p,34}.\newline
 It is shown that type II superstrings  possess hyperbolic
invariance.
 Indeed,  the  four dimensional (4D) ADE supersymmetric quiver gauge theories can be understood
 from  the field description of type II superstrings compactified on three dimensional Calabi-Yau manifolds ($CY_3$)
  with generalized ADE
   geometries associated  with  hyperbolic singularities  \cite{35}.

The aim of this work  is to contribute to these activities  by
investigating  SM quivers involving  Type II  superstring axion
fields  associated with the U(1)$%
_{PQ} $ global symmetry. In the case of standard singularities,  the
Peccei-Quinn realization extends the group symmetry of the SM-Like
model and contributes to the mediation of the electroweak symmetry
breaking. According to the fermion mass scales as well as string
theory data, the allowed axion window is discussed. Then, a
speculation on the size of the underlying internal geometry will be
given. We will extend this discussion to the case of hyperbolic
quivers and we will compare both types of models.

The paper is organized  as follows. In section 2, we build a SM-like
model with a closed string axion field. In section 3, we discuss the
matter coupling terms in intersecting  Type IIA D6-brane model
building. The corresponding  axion window is in the allowed range
$10^{9}GeV\leq f_{\sigma}\leq 10^{12}GeV$ matching with the recent
cosmological results. In section 4. we extend previous results to
the case when an hyperbolic quiver is considered and its modulus
corresponds to the axion using Type IIB D-branes. We study the dependence of
of the closed string
axion decay constant with the string scale in the case when a node
in the quiver is substituted by an hyperbolic one. We find that both scales disantangle allowing in principle to model the $U(1)_{PQ}$ symmetry by the closed string axionic sector. The last section
is devoted to concluding remarks.

\section{Type IIA  stringy constructions of Axion in  quiver models}
At this moment stringy realization of QCD axions has been attempted from several
 points of view. Axion-like particles, scalar CP odd fields
  are ubiquitous in string  theory  compactifications as they appears in effective 4D actions\cite{23,230}
   via the superpartners of geometric moduli fields (also called \textit{saxions}
  ) \cite{290} associated  with  the  string compactification manifolds and also as the the zero modes of $p$-forms
   or in the open string sector as scalars of the D-brane models. They are contained in the
   open and the closed string sector and at present it has became clear that both type of
    axions have to be considered \cite{23} to obtain the axion candidate for QCD axion.

Although axion-like
particles can be  mixed to form
 the physical axion in type IIA construction,  it has been obtained
  that the QCD axion is driven by mainly the open string axion, since the closed string
   axion -which on the other hand is a good candidate to model dark matter- it  becomes the
    longitudinal mode of the massive axions.  Connection with
    standard model physics \cite{38,39,40,401} subject to different constraints on the axion scale \cite{402,403,404} has been also studied in a number of  ways.    \newline

    The string axion depends strongly on the moduli stabilization process to be able to model the QCD axion.  When the moduli stabilization
      process is due to nonperturbative potentials, the saxions and axions get the same
       scale which is either too high or it forces to establish the string scale to be
        at an intermediate scale $10^{12} Gev$ something that with the present bounds
         seems disfavored. According to BICEP2 results the energy inflation scale is at $10^{16}$ Gev and consequently the string scale has  to be higher \cite{281}. The theoretical difficulties of the stringy axions to model out QCD axions, in particular the closed string ones, were pointed out by \cite{240} attending also to the mechanism of moduli stabilization performed in the model.  If the moduli stabilization is performed perturbatively  \cite{241,242,243}
         via D-terms,  then the saxion remains light and
         its perturbative shift symmetry can be broken by non-perturbative effects like
         instanton or gaugino condensation at low scales compatible with the QCD scale,
          while the geometric moduli is stabilized at Msoft scale. Soft mass scale is of the order
           of gravitino mass typically, since they enter in the scalar potential trough
            the Kahler potential. Moreover, it has also been established that  the light
             closed stringy axion in perturbative moduli stabilization decay into Dark
              radiation and they do not acquire an anomalous U(1) masses required to model
              out the QCD axion \cite{371}. With the recent constraints it has become clear that
               the string scale cannot
              be of the order of the axion decay constant, something that generically occurs
              for the closed string axion. This necessary hierarchy between the supersymmetry scale
               and moduli stabilization can  be obtained through the
               the construction of warped throats \cite{fgu}.\newline

    In this paper we will explore a different avenue,  the role of hyperbolic singularities to disentangle both scales. In  the present work,  we consider a toy model in which
     the Peccei-Quinn symmetry is
driven
 by a closed string axion, modeled in an effective field theory whose stringy origin
   could be D6-branes that are wrapped in three cycles before supersymmetry is broken
   completely through some mechanism that we are not interested at the moment.
   However,  this model suffers of the disadvantages already signaled for closed
   string axions. We then perform an slight modification in the model, to change
    the SM-extended quiver into an hyperbolic quiver in order to see
    the role of
     of hyperbolic singularities  in the discussion of axion issue.

More precisely, we build an intersecting D-brane model building. To
start, it is worth to recall that Type IIA D-brane constructions
with orientifold
configurations offer large components to build SM-like field theories \cite%
{15,16,18}. In this framework, a set of three stacks brane model can
accommodate the U(3)$\times $Sp(1) $\times $U(1) gauge symmetry and SM
matter content. A priori, there are many ways to approach the U(1)$_{PQ}$
symmetry. In fact, it can be considered either as a local or as a global
symmetry. In this work, we will enlarge the U(3)$\times $Sp(1) $\times $U(1)
gauge symmetry by incorporating an abelian gauge factor associated with the
U(1)$_{PQ}$ symmetry, which will be considered as a global symmetry. Indeed,
the model will be built form four stacks of intersecting Type IIA D6-brane
model with such a flavor-dependent symmetry distinguishing various fermions
from each other. In this stringy representation, the D6-brane configurations
give the following gauge symmetry
\begin{equation}
\mbox{U(3)}_{a}\times \mbox{Sp(1)}_{b}\times \mbox{U(1)}_{c}\times %
\mbox{U(1) }_{PQ}.  \label{eq1}
\end{equation}
However, the intersections between such D6-branes produce matter
fields. The corresponding physics can be encoded nicely in a graph
which can be easily translated to a quiver diagram. In what follows, the U(1)%
$_{PQ}$ symmetry will be broken by introducing a complex scalar field $\phi
=\rho exp(\frac{i\sigma }{f_{\sigma }})$, where $\sigma $ will be
interpreted as a Type IIA closed string axion. $f_{\sigma }$ is the axion
decay constant. In fact, this complex scalar involves a non vanishing vacuum
expectation value given by $\langle \phi \rangle =f_{\sigma }$. The
Lagrangian associated with this closed string axion field $\sigma $ takes
the following form
\begin{equation}
\mathcal{L}=\partial _{\mu }\sigma \partial ^{\mu }\sigma +\frac{\sigma }{%
f_{\sigma }}F\wedge F,  \label{eq2}
\end{equation}%
where $F$ will be identified with the gauge field strength living on the
D6-brane. It is recalled that the vanishing of the non-abelian anomalies is
implied by the tadpole conditions, while the Green-Schwarz mechanism cancels
the abelian and mixed anomalies. On the perturbative level, the anomalous
U(1)$^{\prime }s$\textbf{\ }get\textbf{\ }massive and survive only as global
symmetries and forbid various couplings to appear. In the resulting
four-dimensional space-time gauge group, the hypercharge is identified with
a massless remained linear combination of the global symmetries.

The SM fermion charges can be set using the vanishing of anomaly
requirements with the presence of flavor constraints related to the
complex scalar field $\phi $. In connection  with higher dimensional
theories, the closed string axion field $\sigma$ can be obtained
either from open string theory or closed string theory. In Type II
superstrings, the closed string axion can be obtained from the NS-NS
or R-R forms wrapping appropriate cycles in the compactified
internal spaces. In the case of Type IIA superstring, the closed
string axion field can be obtained by considering the R-R 3-form
wrapping a 3-cycle $\Sigma $. This scalar field is charged
under the gauge field living on the D6-brane wrapping the same 3-cycle $%
\Sigma $. In this present scenario, we identify this gauge symmetry with the
U(1)$_{PQ}$ symmetry.

The second term of the Lagrangian (\ref{eq2}) can be derived from the Chern
Simon action associated with the same D6-brane world volume. Using string
theory technics, this term takes the following form
\begin{equation}
S\supset \frac{M_{string}^{3}}{4\pi }\int \limits_{R^{1,3}\times
\Sigma }C_{3}\wedge Tr(F\wedge F),  \label{eq3}
\end{equation}
where $C_3$ is the R-R 3-form appearing in Type IIA superstring.
String theory compactification can be explored to produce the
complex filed $\phi $ charged under the U(1)$_{PQ}$ symmetry. In
connection with four dimensional theories, the field generates
allowed invariant coupling terms. In fact, there are many ways to
implement   such interaction coupling terms. Indeed, it is possible
to introduce a general form, into SM-like models, given by
\begin{equation}
y_{f}\left( \frac{\phi }{M_{s}}\right) ^{n_{f}}hf\overline{f},  \label{eq4}
\end{equation}%
where $f$ refers to the SM fermion fields. The coupling parameters $y_{f}$
correspond to the Yukawa constants. $M_{s}$ is the string scale mass and $%
n_{f}$ is a specific number which will be determined later by exploring SM
and string theory data. Throughout the present work, a particular attention
will be on such a number and its relation to the physics of the closed
string axion in SM-like models. More precisely, the discussion will depend
on the possible values of $n_{f}$. Using the global U(1)$_{PQ}$ symmetry
acting on the fields as follows
\begin{equation}
\phi \rightarrow e^{iq_{\phi }\alpha }\phi ,\quad h\rightarrow
e^{iq_{h}\alpha }h,\quad f\rightarrow e^{iq_{f}\alpha }f,  \label{eq5}
\end{equation}%
one can get some constraints on $n_{f}$ and gauge charges. In fact, the
invariance of the equation (\ref{eq4}) produces the following equation
\begin{equation}
n_{f}=-\frac{q_{h}+q_{f}+q_{\overline{f}}}{q_{\phi }}.  \label{eq6}
\end{equation}%
Exploring SM fields, one can solve this equation using intersecting
D6-brane charge assignments. The latter depend on the considered
Type IIA string compactification. Indeed, the model proposed here is
based on a particular choice of intersecting numbers of 3-cycles of
the middle cohomology inspired by intersecting D6-brane model
building obtained from the  toroidal orientifold compactification.
To obtain the above gauge symmetry from string theory, we can
consider four stacks of D6-branes indicated by
D$6_{a,b,c,PQ}$-branes. The intersecting number $I_{\alpha \beta }$
$=\Sigma _{\alpha }\circ \Sigma _{\beta }$ between the D$6_{\alpha
=a,b,c,PQ}$-brane and D$6_{\beta =a,b,c,PQ}$-brane is usually given
in terms of the 3-cycles $\Sigma _{\alpha ,\beta }$ intersections.
In the Type IIA superstring theory compactified on a factorized six
dimensional tori $T^{6}=T_{1}^{2}\times T_{2}^{2}\times T_{3}^{2}$,
the number $I_{\alpha \beta }$ is given  by winding numbers. Roughly
speaking, the  first  model that we present here requires D6-brane
configurations with the following intersection numbers

\begin{eqnarray}
I_{D_{a}D_{b}} &=&3, \quad I_{D_{a}D_{c}}=-2,\quad I_{D_{a}D_{c^{\ast }}}=-1,
\notag \\
I_{D_{a}D_{PQ}} &=&-1,\quad I_{D_{a}D_{PQ^{\ast }}}=-2,\quad
I_{D_{PQ}D_{b}}=3,  \label{eq7} \\
I_{D_{PQ}D_{c}^{\ast }} &=&-3.  \notag
\end{eqnarray}%
The other numbers are set to zero. These intersection numbers
generate the resulting spectrum and the associated gauge symmetry
charges. The gauge fields, leptons, quarks and scalar fields can be
encoded nicely in a graph  called a quiver as illustrated in figure 1.

\begin{center}
\begin{figure}[th]
\begin{center}
{\includegraphics[width=10cm]{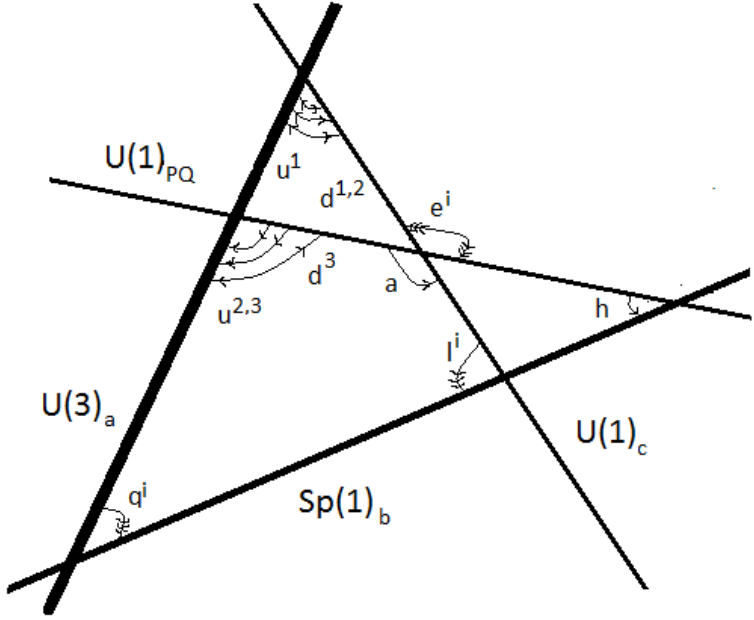}} \vspace*{-.2cm}
\end{center}
\caption{Graph of the intersecting D6-brane model.}
\label{fig1:fig2x}
\end{figure}
\end{center}

In this graph, the lines
represent the gauge fields and the intersection points give the
matter fields. This stringy model, which is based on the above
intersecting D6-branes, is governed by a particular choice of the
charges which are listed in the table 1.

\begin{center}
\begin{tabular}{|c|c|c|c|c|c|c|c|c|c|}
\hline
Fields & $q_{L}^{i}$ & $\overline{u}_{R}^{1}$ & $\overline{d}_{R}^{1,2}$ & $%
\overline{u}_{R}^{2,3}$ & $\overline{d}_{R}^{3}$ & $l_{L}^{i}$ & $\overline{e%
}_{R}^{i}$ & $h$ & $\phi$ \\ \hline
$\mbox{U}(1)_{c}$ & 0 & 0 & 0 & 1 & -1 & 0 & -1 & 1 & 1 \\ \hline
$\mbox{U}(1)_{PQ}$ & 0 & 1 & -1 & 0 & 0 & 1 & -1 & 0 & -1 \\ \hline
$\mbox{U}(1)_{Y}$ & 1/6 & -2/3 & 1/3 & -2/3 & 1/3 & -1/2 & 1 & -1/2 & 0 \\
\hline
\end{tabular}

\bigskip Table1:The field content corresponding to the Hypercharge
combination $Y=\frac{1}{6}Q_{a}-\frac{1}{2}Q_{c}-\frac{1}{2}Q_{PQ}$. The
index $i(=1,2,3)$ denotes the family index.
\end{center}

\section{Matter coupling terms}

Having built the intersecting D-brane geometry, we discuss now the
corresponding coupling terms. In this stringy representation, the
fermions are distinguished by their  U(1)$_{c,PQ}$ charges. The
latter will forbid the renormalisable Yukawa couplings for the
U(1)$_{PQ}$ charged quarks, but would allow them via effective
highly suppressed couplings through the complex field $\phi $
carrying appropriate charges under U(1)$_{PQ}$. Upon spontaneous
breaking of the U(1)$_{PQ}$ symmetry by the vev of $\phi $, the
effective Yukawa couplings appearing in  (\ref{eq4}) can be
generated with positive numbers $n_{f}$ solving the above charge
equations. Using the intersection numbers presented above, one can
determine the low-energy effective Yukawa lagrangian. A close
inspection shows that the allowed coupling terms read as
\begin{eqnarray}
\zeta _{Yuk} &=&y_{u^{2}}h^{\dagger }q_{L}\overline{u}_{R}^{2}+y_{u^{3}}h^{%
\dagger }q_{L}\overline{u}_{R}^{3}+y_{d^{3}}hq_{L}\overline{d}_{R}^{3}+\
y_{e^{i}}hl_{L}^{i}\overline{e}_{R}^{i}+y_{u^{1}}\frac{\phi }{M_{s}}%
h^{\dagger }q\overline{u}_{R}^{1}  \notag \\
&+&y_{d^{1}}\frac{\phi ^{\dagger }}{M_{s}}hq\overline{d}_{R}^{1}+y_{d^{2}}%
\frac{\phi ^{\dagger }}{M_{s}}hq\overline{d}_{R}^{2}+y_{v^{i}}\frac{\phi ^{2}%
}{M_{s}^{3}}h^{\dagger }h^{\dagger }l_{L}^{i}l_{L}^{i}+h.c.,  \label{eq8}
\end{eqnarray}%
where the heavy fermions are directly connected with the electroweak
symmetry breaking sector and they have four-dimensional Yukawa couplings.
The lighter ones are related with the electroweak symmetry breaking sector
via the complex field $\phi $ and they have high dimension suppressed
Yukawa-like couplings involving powers of $\left( \phi /M_{s}\right) $ or $%
\left( \phi ^{\dagger }/M_{s}\right) $. As mentioned before, the powers are
required by the flavor charges and some physical data of SM. Using the
following approximation
\begin{equation}
\phi ^{2}h/M_{s}^{3}\sim \left( \phi /M_{s}\right) ^{3},  \label{eq9}
\end{equation}%
we can show that $n_{f}$ should take the following specific values
\begin{equation}
n_{f}=0,1,3.  \label{eq10}
\end{equation}%
It is remarked that $n_{f}=0$ corresponds to the absence of the closed
string axion field. The stringy origin of the axion field can be explored to
give more information on $n_{f}$. This could bring a string theory
interpretation of such numbers. This discussion is beyond the scope of the
present work, however, we will try to address this question elsewhere.

\bigskip Having fixed the values of $n_{f}$, we discuss now the electroweak
symmetry breaking associated with the lagrangian (\ref{eq8}). After such a
breaking symmetry with $\left \langle h\right \rangle =v$, and using a
simple scaling, the resulting lagrangian for the mass coupling term takes
the form
\begin{equation}
\zeta _{mass}=y_{f}\epsilon vf\overline{f}.  \label{eq11}
\end{equation}%
Forgetting about the fermion mass hierarchy which could be encoded in the
corresponding Yukawa constants $y_{f}$, $\epsilon $ is a suppression factor
related to the closed string axion decay constant $f_{a}$ and the $M_{s}$
via the relation
\begin{equation}
f_{\sigma }=\epsilon ^{\frac{1}{n_{f}}}M_{s}.  \label{eq12}
\end{equation}%
This reveals that the decay constant of the closed string axion is
proportional to the mass string scale. Breaking the electroweak symmetry at
the usual scale $v\simeq $ $10^{2}GeV$, and using the ratio of certain
suitable fermion masses, we can estimate the suppression factor $\epsilon$.
In fact, it is given by
\begin{equation}
\epsilon \simeq \frac{m_{s}}{m_{t}}\simeq 10^{-3},  \label{eq13}
\end{equation}%
where $m_{s}$ and $m_{t}$ are the strange and the top quark masses.

According to this numerical data and taking $M_{s}=10^{12}GeV$, the
so-called axion window is obtained for the above mentioned values of
$n_{f}$.  Using the above requirements, it is found the following
energy scales

\begin{center}
\begin{tabular}{|c|c|c|}
\hline
$n_{f}$ & 1 & 3\\ \hline
$f_{\sigma}\left( GeV\right) $ & 10$^{9}$ & 10$^{11}$ \\ \hline
\end{tabular}
\bigskip

Table2: The possible axion decay magnitudes.
\end{center}

The computation shows that
\begin{equation}
10^{9}GeV\leq f_{\sigma }\leq 10^{12}GeV,  \label{eq14}
\end{equation}%
giving a good argument with the astrophysical observations and the
cosmological considerations made on the axion decay constant region \cite%
{22,23}.

In connection with the large volume scenario, the closed string axion field
can be used to estimate the size of the compact 3-cycle in which D6-brane
can be wrapped. In type IIA superstring, the volume of the six-dimensional
internal space $V_{6}$ and the volume of the compact three-cycle $V_{3}$ on
which the D6-brane is wrapped are related to the axion decay constant as
reported in \cite{22,220,23}. In the present study, this relation can be
rewritten as follows
\begin{equation}
V_{6}=128\pi \epsilon ^{\frac{2}{n_{f}}}V_{3}^{2}.  \label{eq15}
\end{equation}%
It is observed that $n_{f}=1$ matches with the large volume wrapped geometry
scenario required by $V_{6}\sim V_{3}^{2}$.

To make a contact with supersymmetric quiver theories as well as
their type IIA string interpretations, one may ask a basic question
regarding the volume of three cycle, in which a D6-brane is wrapped,
associated with the Peccei-Quinn symmetry U(1)$_{PQ}$. In fact, this
volume is linked with the corresponding coupling gauge symmetry
\cite{36}. In the breaking symmetry, the volume becomes very large.
This might be not a problem in type IIA super string since one
considers local geometries  known as local type II geometries
\cite{36}. This observation should be addressed in elsewhere.

\section{Peccei-Quinn axion from the hyperbolic SM quiver in Type IIB superstring}
In this section, we investigate Peccei-Quinn axion from the
hyperbolic SM quiver.  Indeed, the formulation shown in the previous
section faces two difficulties, on one hand it has no supersymmetry which is a handicap in order  to find a stringy realization, on the other hand the axion decay constant is proportional to the mass scale and to lie int eh cosmological acceptable window of values it requires the string mass scale to be intermediate which seems to be observationally disfavored. In order to discuss a possible improvement of this situation just taking into account the closed string sector we are going to study the quiver realization of the Peccei-Quinn symmetry in an hyperbolic quiver.
  To deal with the non supersymmetric embedding in String theory seems quite natural to think in hyperbolic spaces,  since in general hyperbolic singularities are often associated  with  de Sitter
  vacuum with no supersymmetry left at all. Consequently, we can look for hyperbolic SM extensions
   realized in terms of D-brane configurations on non-supersymmetric hyperbolic singularities.
   There can be however, other models as those discussed in \cite{313} with some supersymmetry
    left which can be also of interest to study. The extension we are going to discuss in this work consists in subtituting one of the 2-cycles of the comapctification manifold by an hyperbolic 2-cycle. The model we will obtain is $N=2$ supersymmetric gauge theory in 4D.
     Type II superstrings  possess a not so well-explored hyperbolic invariance. This invariance
      appears as a result of having D-branes compactified on manifolds with hyperbolic singularities.
       Also in the literature, it  has been explored the role of S-branes realized in terms
        of NS5-branes and they naturally appear in the context of hyperbolic
        geometry.\\
         It recalled  that many  string  quivers have been built  from non trivial
         singularities using Dynkin  graphs.  In this way, the
physical content of such  quivers  can be encoded in   Dynkin
diagrams using root systems. One may have three models   classified
as follows:
\begin{enumerate}
\item ordinary ADE quiver gauge theories
\item  affine ADE quiver gauge theories
 \item indefinite quiver gauge theories.
 \end{enumerate}

In string theory, the first  class  can be  elaborated  from
D-branes wrapping a collection of intersecting  cycles according to
ordinary ADE Dynkin diagrams of finite Lie algebras. The second
class concerns conformal gauge theories obtained from  D-branes
wrapping cycles  used in the deformation of  elliptic singularities.
However,   the  last   class  is the  more complicated one. They are
only few models  examining     the hyperbolic subset which will be
reconsidered here to discuss the axion field. These models can be
obtained by using the same technics explored in the construction  of
the Kac-Moody Dynkin diagrams from the finite ones by adding an
affine node.
 In   the   elaboration of  quivers,  it has been used
  a nice correspondence between the roots  and two-cycles  on which
  D-branes are wrapped.  More specifically, to each simple
  root $\alpha$ we associate a single node  encoding a gauge
  factor.  Roughly speaking,  it has been realized that the  hyperbolic
    root systems  can be obtained from  affine
ones.  Indeed,  consider a   affine simply laced Kac-Moody algebra
with rank $r$. We denote  its   simple roots by  $\alpha _{i},\;
i=0,1,...,r$.  One can construct a root associated with affine
extension called  affine root given by
 \begin{equation} \alpha _{0}=\delta
-\sum_{i=1}^{r}d_{i}\alpha _{i}
\end{equation} where $\delta $  is
the usual imaginary root and  $\alpha _{i}$ are the simple real
ones. To obtain  a hyperbolic root system, one should add  an extra
simple root  noted $\alpha _{-1}$ satisfying the following  Cartan
intersection numbers \bea \alpha _{-1}.\alpha _{-1}&=&2 \nonumber\\
\alpha _{-1}\alpha _{0}&=&-1\\ \alpha _{-1}\alpha _{i}&=&0
\nonumber. \eea For simply laced algebras with symmetric Cartan
matrices, it is recalled that
\begin{equation}
\alpha _{-1}=\gamma -\delta
\end{equation}%
where  $\gamma $ is a second basic imaginary root being related to
$\delta $ by
\begin{equation}
 \gamma ^{2}-2\gamma .\delta +\delta ^{2}=2,
\end{equation}
 It is easy to show that
\begin{eqnarray}
{\gamma }^{2} &=&{\delta }^{2}=0, \qquad {\gamma }{\delta }=-1,  \notag  \nonumber \\
{\gamma .}\alpha _{0} &=&1;\qquad {\gamma .}\alpha _{i}=0,\ i>0.
\end{eqnarray}
It follows that hyperbolic Lie algebras  involves  two basic imaginary roots $%
\gamma $ and $\delta $  given   by \ba \delta
&=&\sum_{i=0}^{r}d_{i}\alpha _{i} \nonumber\\ \gamma
&=&\sum_{i=-1}^{r}d_{i}\alpha _{i},  \label{torr} \ea where $d_{i}$
($0\leq i\leq r$) are the usual Dynkin weights. It is noted that   $
d_0=d_{-1}=1$. As in the case of Kac-Moody algebras, these roots
generate two elliptic curves
\begin{equation}
T_{+}^{2}=\sum_{i=0}^{r}d_{i}\left[{\bf CP}_{i}^{1}\right];\qquad
T_{-}^{2}=\sum_{i=-1}^{r}d_{i}\left[{\bf CP}_{i}^{1}\right],
\label{tor}
\end{equation}
where ${\bf CP}_{i}^{1}$ form  a basis of $H_{2}(K3,\mathbb{Z})$ in
II superstrings on CY3s with deformed ADE singularities.
 These two torus  are related to  two couplings $g_{\pm }$  respectively
\begin{equation}
g_{-}^{-1}=\sum_{i=0}^{r}d_{i}g_{i}^{-2};\qquad
g_{+}^{-1}=\sum_{i=-1}^{r}d_{i}g_{i}^{-2};
\end{equation}
where $g_{-}^{-1}$  and $g_{+}^{-1}$ are  the gauge coupling of the gauge
group engineered on the hyperbolic  and affine  nodes  respectively.
From Type IIB superstring theory point of view, the parameters
$g_{\pm }$  are given by
 \ba
g_-^{-1}= exp(-\phi)-\chi \nonumber\\
 g_+^{-1}=exp(-\phi)+\chi
 \ea
 where $\phi$   and $\chi$ are the dilaton and the axion fields of  ten dimensional
 type IIB spectrum. They are connected  by the following relation
\begin{equation}
g_{+}^{-1}=(g_{-1})^{-2}+g_{-}^{-1},
\end{equation}%
Replacing  $g_{+}^{-1}$ and $g_{-}^{-1}$ by the dilaton and the
axion expressions,  one can obtain  the gauge coupling  associated to the hyperbolic node
\be (g_{-1})^{-2}=2\chi. \ee

 There exist  holomorphic $(2,0)$
    nontrivial two-cycles such that one can define  the holomorphic volume of the two
     cycles $\xi$ as shown in \cite{35}
\begin{equation}
\xi_j=\int_{{C}_j}\Omega^{(2,0)}
\end{equation}
The volume $v_{-1}$ of the hyperbolic 2-cycle $C_{-1}$ is given by the vev of
the axion modulus $\chi$.

Moreover demanding  the  positivity of the $g_{\pm}$ couplings
constants oblige the range of $\phi$ and $\chi$ moduli to satisfy a
constraint. Taking the dilaton as the basic modulus with an
arbitrary real value, the range of the axion RR field has an upper
bound given by, \be\vert \chi\vert\le\phi.\ee

This
equation shows  that the RR axion field vacuum expectation value can be identified with the
volume of the cycle associated with  hyperbolic quiver. Since the axion decay function corresponds to $f_{\sigma}=<\chi>$ then for the case of the hyperbolic quiver this scale can be disantangled from the $M_s$ and fixed independently such that it will satisfy the allowed axion decay window at the same time that we can estimate $M_s\simeq 10^{18} Gev$ by properly adjusting the vev of the dilaton $\phi$ defined also in terms of the gauge coupling of the hyperbolic and affine nodes $$g_s^{-1}=\frac{1}{2}[(g_{-})^{-1}+(g_{+})^{-1}].$$

 For non-zero axion, there is a $U(N_{-1})$ gauge symmetry with non-zero
 coupling $g$ on the hyperbolic node of the underlying hyperbolic
 node, which will be identified with the Peccei-Quinn symmetry U(1)$_{PQ}$.

  It is
recalled  that   the D-brane realization of hyperbolic ADE gauge
theories can be obtained using D3 and D5-branes. Indeed,   we need
two stuck of $N_{-1}$  and $N_{0}$ coincident regular D3-branes and
$N=\sum_{i=1}^{r}N_{i}$ fractional D5-branes  of  Type IIB
superstring. Each block of the $r$ sets of $N_{i}$ D5-branes is
wrapped over two cycles of the deformed ADE geometry of the local
CY3. For  an arbitrary Lie algebra with rank $r$, the corresponding
quiver model has  the following  gauge group
\begin{equation}
G =U\left( N_{-1}\right) \times U\left( N_{0}+d_{0}N_{-1}\right)
\times \prod_{i=1}^{r}U\left( N_{i}+d_{i}\left( N_{-1}+N_{0}\right)
\right).
\end{equation}%
The above quiver studied in the previous section can be recovered by
\be
 N_{-1}=N_0=N_1=N_2=1.
 \ee
It produces a   quiver model where  the Peccei-Quinn symmetry
U(1)$_{PQ}$ is  associated with  the  hyperbolic node.

\section{Conclusions}

In this work, we have discussed  the stringy axion from quiver
embedded in type II superstrings. First,
 we have presented a stringy axion extension of SM obtained
from a set of four stacks of intersecting Type IIA D6-branes in the
presence of the Peccei-Quinn symmetry U(1)$_{PQ}$. We have
introduced a complex scalar field $\phi =\rho
exp(\frac{i\sigma}{f_{\sigma}})$, where $\sigma$ is the closed
string axion, generating a general fermion Yukawa coupling weighted
by a flavor-dependent power $n_{f}$. It has been found that the
corresponding axion scale is in the allowed range $10^{9}GeV\leq
f_{\sigma}\leq 10^{12}GeV$. This discussion has been based on some
possible values of $n_{f}$ given by 1 and 3. A fast reflection on
Type IIA superstring spectrum shows that these values could have an
interpretation in terms of the order of R-R antisymmetric tensor
fields. Then, we have investigated the axion field form hyperbolic
quiver in type IIB superstring. We have build  quiver model with the
Peccei-Quinn symmetry U(1)$_{PQ}$ associated with hyperbolic node.
We  have obtained  that the axion decay constant is directly
associated to the volume of the hyperbolic node of the quiver. This
preliminary study indicates that string scale may be disentangled
from the axion decay constant window for the case of hyperbolic
compactifications. A proper global string compactification is needed in order to fully realize this idea. If our hypothesis is correct in global compactifications this fact can provide a way realize Peccei-Quinn symmetry by means of closed string axions. Hyperbolic
geometries have less number of moduli than usual compactifications which can be also an advantage for moduli stabilization processes.  Indeed for hyperbolic spaces with dimension larger or equal than three the
Mostow rigidity theorem applies and it simplifies the moduli
stabilization analysis, since there are only two significative
length scales parametrizing the manifold. With regard to the moduli
stabilization we are considering that the saxions can be stabilized
through fluxes acting perturbatively together with the
Fayet-Illiopolous terms on the D-term potentials in the spirit of
\cite{241,242,243} leaving the axions light.

 The toy model considered here is not supersymmetric. Usual embedding
  in intersecting brane models must contain some supersymmetry,  also  the hyperbolic quiver we
  are proposing preserves $N=2$ supersymmetry in 4D and consequently the model should be extended.
 However, there are other hyperbolic models that break all supersymmetries and in that
   sense they can be more natural uplifts for non supersymmetric effective actions. That
   is for example the case of compact target spaces that contain 2-dimensional hyperbolic
    spaces \cite{313, erika} \footnote{Though it does not automatically means that these models are unstable.}.

This work comes up with many questions related to possible extended field models. A concrete
 one concerns the link with string theory compactification
involving various closed string axion fields. It is recalled, in passing,
that open string axions used in consistent D-brane constructions are the ones that contribute mostly
 to the QCD axion. Their role should be
explored in the building of such extended theories.  The Higgs
sector also has a nontrivial interactions
 in the scalar potential with the axion one  that we have not considered. It will be of
interest to investigate all of these aspects in the future and make contact with the
closed string sector in standard and hyperbolic compactifications.

\section{Acknowledgements} The authors would like to deeply thank to A. Uranga and W. Staessens
 for helpful comments. MPGM is supported by Mecesup ANT1398, Universidad de Antofagasta, Antofagasta
1121103 (Chile).

\end{document}